\documentclass[prb,aps,twocolumn,showpacs,floatfix]{revtex4-1}
\usepackage{amsfonts}
\usepackage{amssymb}
\usepackage{hyperref}
\usepackage{graphicx}
\usepackage{dcolumn}
\usepackage{bm,amsmath,verbatim}
\usepackage{mathrsfs}
\usepackage{color}

\hypersetup{colorlinks,
linkcolor=blue,          citecolor=blue,        filecolor=blue,      urlcolor=blue           }

\def\be{\begin{equation}}
\def\ee{\end{equation}}
\def\bea{\begin{eqnarray}}
\def\eea{\end{eqnarray}}
\def\bse{\begin{subequations}}
\def\ese{\end{subequations}}

\def\be{\begin{eqnarray}}
\def\ee{\end{eqnarray}}

\begin{document}

\title{Unconventional quantum Hall effects in two-dimensional massive spin-1 fermion systems}
\author{Yong Xu}
\email{yongxuph@umich.edu}
\author{L.-M. Duan}
\affiliation{Department of Physics, University of Michigan, Ann Arbor,
Michigan 48109, USA}
\affiliation{Center for Quantum Information, IIIS, Tsinghua University,
Beijing 100084, PR China}

\begin{abstract}
Unconventional fermions with high degeneracies in three dimensions beyond Weyl and Dirac fermions
have sparked tremendous interest in condensed matter physics.
Here, we study quantum Hall effects (QHEs) in a two-dimensional (2D) unconventional fermion system
with a pair of gapped spin-1 fermions. We find that
the original unlimited number of zero energy Landau levels (LLs) in the gapless case develop into a series of bands, leading to a novel QHE phenomenon that the Hall conductance
first decreases (or increases) to zero
and then revives as an infinite ladder of fine staircase when the Fermi surface is moved toward zero energy, and
it suddenly reverses with its sign being flipped due to a Van Hove singularity when the Fermi surface is moved across zero.
We further investigate the peculiar QHEs in a dice model with a pair of spin-1 fermions, which agree well with
the results of the continuous model.
\end{abstract}
\maketitle

\section{Introduction}

\begin{figure*}[t]
\includegraphics[width=\textwidth]{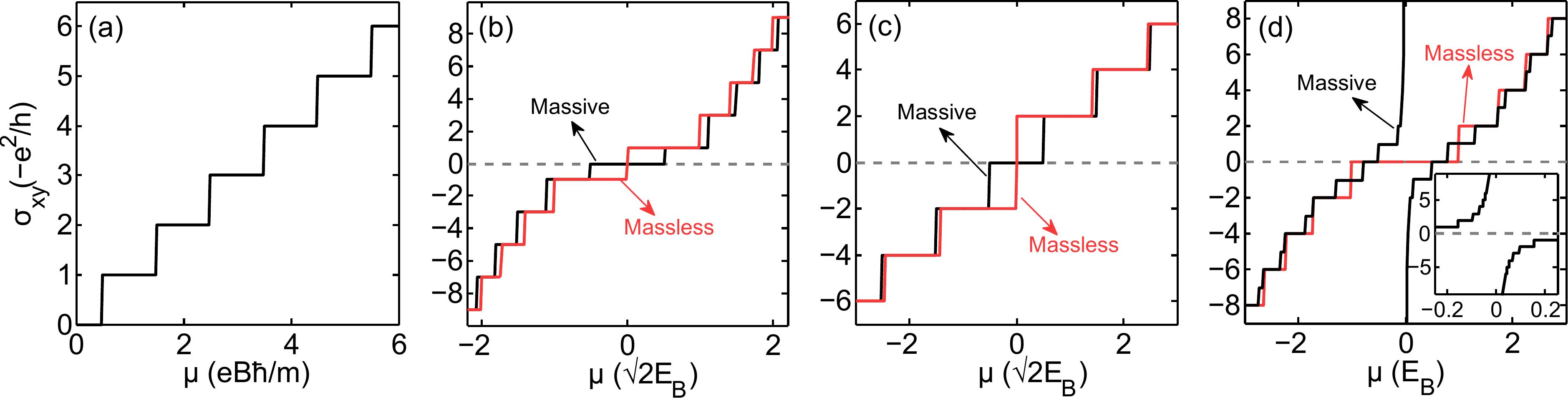}
\caption{(Color online) The Hall conductance with respect to the chemical potential in a conventional 2D electron gas (a),
where $B$ and $m$ denote the magnetic field strength and the electron mass, respectively; in a single
layer graphene (b) and bilayer graphene (c); in a system with two triply degenerate points (d) with a zoomed in
view in the inset.
In (b), (c) and (d), the red and black lines represent the massless and massive cases with the mass $m_z=0$ and $m_z=0.5E_B$,
respectively. Here, $E_B=v\sqrt{eB\hbar}$ with $v$ denoting the electron velocity.}
\label{Fig1}
\end{figure*}

One of the most important and intriguing phenomena in condensed matter physics is
the quantum Hall effect, the phenomenon that the Hall conductance becomes quantized
(i.e., $\sigma_{xy}=-ne^2/h$ with $n$ being integer and $e$ being the electric charge)
at low temperature in 2D electron gases subject to strong magnetic fields. This effect
is originated from the formation of LLs with a quantized Thouless-Kohmoto-Nightingale-den Nijs
(TKNN) number~\cite{TKNN1982} in these systems with magnetic fields. Each occupied LL contributes a $-e^2/h$ Hall conductance and if no LLs
are occupied, the Hall conductance completely vanishes as shown in Fig.~\ref{Fig1}(a). Apart from the conventional QHE
in 2D electron gases with parabolic dispersion, the unconventional QHE in 2D
Dirac materials with relativistic dispersion such
as a single layer graphene and bilayer graphene was discovered that the Hall
conductance can take only odd or even numbers~\cite{Gusynin2005,Neto2006,Firsov2005,Kim2005,Novoselov2006,McCann2006}
(without considering the spin freedom)
as shown in Fig.~\ref{Fig1}(b) and (c).
In addition, these materials always exhibit the nonzero QHE plateaus because of the
existence of zero energy LLs unless these zero energy levels are gapped and
neither particle nor hole LLs are occupied.

Other than 2D, the study of relativistic fermions such as Dirac and Weyl fermions
in three-dimensional systems, a counterpart of Dirac fermions in graphene, has
seen a rapid progress~\cite{Wan2011prb,Teo2012,
Neupane2014,Cava2014,ZKLiu2014,Lu2015,Xu2015,Lv2015,Armitage2017}. Beyond these fermions permitted in particle physics, a new fermion violating
Lorentz invariance~\cite{Yong2015,Soluyanov2015Nature}, which is called type-II Weyl fermions~\cite{Soluyanov2015Nature} (also called structured Weyl fermions~\cite{Yong2015}),
has been discovered in superfluids~\cite{Yong2015} and condensed matter materials~\cite{Soluyanov2015Nature,Bergholtz2015PRL,Deng2016,Kaminskyi2016}. Recently, another type of new fermions in three dimensions
beyond conventional fermions in particle physics were predicted in solid-state materials~\cite{Kane2016,Bernevig2016,Weng2016PRB,Hasan2016,Wan2016,Ding2016,Felser2017};
they are named unconventional fermions with highly degenerate points that are described by an effective Hamiltonian $H={\bm k}\cdot\hat{{\bm S}}$ with
${\bm k}$ being the momenta and $\hat{\bm S}$ being the angular momentum matrices.
These highly degenerate fermions have also been studied in 2D systems~\cite{Sutherland1986,Vidal1998,Rizzi2006,Bercioux2009,YingRan2012,Roessner2011,Raoux2014,Raoux2014,Biswas2016,Kovac2017,
Watanabe2011,Lan2011,Nicol2014,YangGao2015,Danwei2016,
Dagotto1986,Xing2010,Manninen2010,Goldman2011,
Chamon2010,Shengyuan2016}, especially for spin-1
fermions, which can be realized in the dice model~\cite{Sutherland1986,Vidal1998,Rizzi2006,Bercioux2009,YingRan2012,Roessner2011,Raoux2014,Raoux2014,Biswas2016,Kovac2017},
Lieb model~\cite{Dagotto1986,Xing2010,Manninen2010,Goldman2011}, Kagome model~\cite{Chamon2010} and solid-state materials~\cite{Shengyuan2016}.
For massless spin-1 fermion systems without gaps, QHEs have been
investigated and it has been found that each spin-1 point contributes an integer Hall conductance~\cite{Lan2011}. The conductance vanishes when the Fermi
surface lies in the first gap of Landau levels (LLs) (the gap between the
zero energy and the first nonzero energy levels) despite the existence of an unlimited number of zero energy LLs. For the dice
model where there exist a pair of spin-1 points, the Hall conductance takes even numbers, reminiscent of the bilayer grephene~\cite{Biswas2016}.

In this paper, we study the QHE in a 2D system with a pair of spin-1 fermions with a mass term $m_z\hat{S}_z$. By exploring a continuous model
under magnetic fields, we find that the infinite degeneracy of zero energy LLs in the massless case is
lifted in the massive case, and these LLs develop into a series of nonzero energy levels. Interestingly, these
levels also contribute to the Hall conductance as the Fermi surface lies in the gap among these levels,
leading to a novel QHE that the Hall conductance revives in the opposite direction as an infinite ladder of fine staircase
when the Fermi surface is moved into the fine structure developed from the original zero energy bands and then experience a sudden
reversal with its sign being flipped due to a Van Hove singularity when the Fermi surface is moved across zero energy.
We further study the QHE in the dice model with a pair of massive spin-1 fermions by calculating its LLs,
Chern numbers and chiral edge states. All these results are in good agreement with the continuous model's ones.

\section{Continuous model}

We start by considering the following 2D continuous model with a single spin-1 point
\begin{equation}
H=v\hat{p}_x\hat{S}_x+v\hat{p}_y\hat{S}_y+m_z\hat{S}_z,
\label{H1}
\end{equation}
where $\hat{p}_\nu=-i\hbar\partial_\nu$ with $\nu=x,y$ are momentum operators,
$v$ is a real parameter, $m_z$ is the mass term that can open a gap, and
$\hat{S}_\nu$ with $\nu=x,y,z$ denote the angular momentum matrices for spin 1,
given by
\begin{gather}
\hat{S}_x=\frac{1}{\sqrt{2}}\left(
       \begin{array}{ccc}
         0 & 1 & 0 \\
         1 & 0 & 1 \\
         0 & 1 & 0 \\
       \end{array}
     \right),
\hat{S}_y=\frac{1}{\sqrt{2}}\left(
       \begin{array}{ccc}
         0 & -i & 0 \\
         i & 0 & -i \\
         0 & i & 0 \\
       \end{array}
     \right), \nonumber \\
\hat{S}_z=\left(
       \begin{array}{ccc}
         1 & 0 & 0 \\
         0 & 0 & 0 \\
         0 & 0 & -1 \\
       \end{array}
     \right),
\end{gather}
which satisfy the angular momentum commutation rule.

The eigenenergy of this Hamiltonian is $E({\bm k})=0,\pm\sqrt{v^2 k_x^2+v^2 k_y^2+m_z^2}$
in the momentum space $(k_x,k_y)$ with a flat zero energy band. It is evident that,
when $m_z=0$, the system exhibits a triply degenerate point at $k_x=k_y=0$
as shown in Fig.2 (a), whose degeneracy can be lifted by $m_z$ as shown in Fig.2 (b).

Since we will focus on the dice model that possesses two triply degenerate points for
a concrete realization, we consider the following Hamiltonian with two degenerate points
\begin{equation}
H=\tau_z(v\hat{p}_x\hat{S}_x+v\hat{p}_y\hat{S}_y+m_z\hat{S}_z),
\end{equation}
where $\tau_z$ is a Pauli matrix and denotes two triply degenerate points. In the presence of magnetic fields
along $z$ described by a vector potential ${\bf A}$,
we replace the momentum operators $\hat{p}_\nu$ with $\nu=x,y$ with the generalized
momentum operators $\hat{\pi}_\nu=\hat{p}_\nu+eA_\nu$ and write down the Hamiltonian in the following form,
\begin{equation}
H_M=\tau_z(v\hat{\pi}_x\hat{S}_x+v\hat{\pi}_y\hat{S}_y+m_z\hat{S}_z).
\end{equation}
To solve the LLs, we
define $\hat{a}=(\hat{\pi}_x-i\hat{\pi}_y)v/(\sqrt{2}E_B)$ which satisfies $[\hat{a},\hat{a}^\dagger]=1$ and recast the Hamiltonian
into the form
\begin{equation}
H_M=\tau_z\left(
  \begin{array}{ccc}
    m_z & \hat{a}E_B & 0 \\
    \hat{a}^\dagger E_B & 0 & \hat{a} E_B \\
    0 & \hat{a}^\dagger E_B & -m_z \\
  \end{array}
\right),
\label{HM}
\end{equation}
where $E_B=v\sqrt{eB\hbar}$ with $B$ being the magnetic field strength.
Its eigenstates can be written as $\Psi=\left(
                                       \begin{array}{ccc}
                                         a|m-1\rangle & b|m\rangle & c|m+1\rangle \\
                                       \end{array}
                                     \right)^T
$ with $a$, $b$ and $c$ being the parameters satisfying the normalization condition.
For clarity, let us first choose $\tau_z=1$ and focus on a single degenerate point.
When $m=-1$, we have $a=b=0$ and $c=1$ with energy being $-m_z$; when $m=0$, we have
$a=0$ with energy being $(-m_z\pm\sqrt{m_z^2+4E_B^2})/2$; when $m>0$, the Hamiltonian is given by
\begin{equation}
H(m)=\left(
    \begin{array}{ccc}
      m_z & E_B\sqrt{m} & 0 \\
      E_B\sqrt{m} & 0 & E_B\sqrt{m+1} \\
      0 & E_B\sqrt{m+1} & -m_z \\
    \end{array}
  \right).
\end{equation}

\begin{figure}[t]
\includegraphics[width=3.2in]{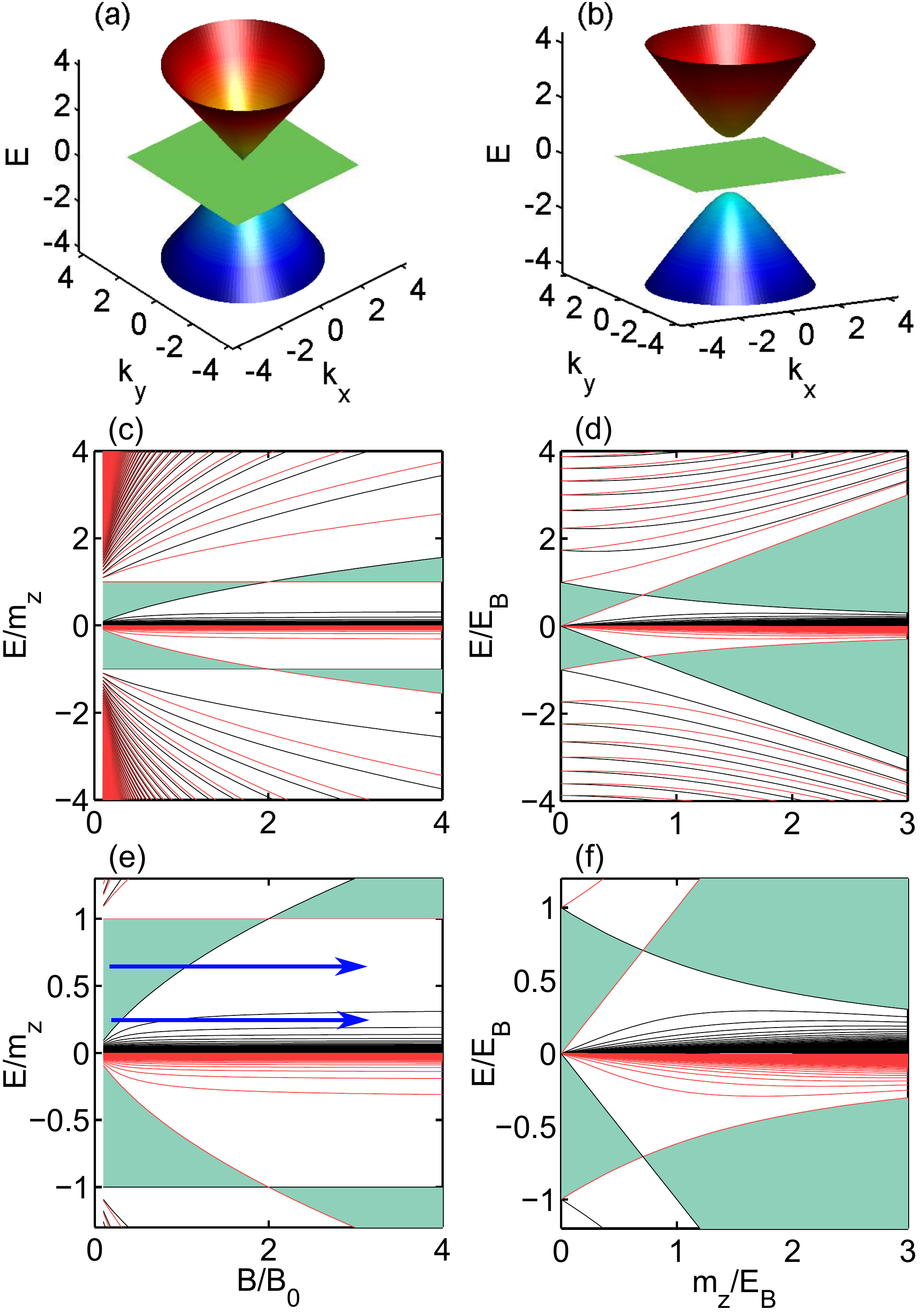}
\caption{(Color online) Eigenenergy of the Hamiltonian (\ref{H1}) for $m_z=0$ in (a)
and $m_z>0$ in (b) with zero energy states denoted by the green plane. Energy of
LLs as a function of $B$ for fixed $m_z$ (this $m_z$ is taken as the energy unit) in (c) and $m_z$ for fixed $B$ (with the
corresponding $E_B$ taken as the energy unit) in (d).
In (c) and (d), the black and red lines correspond to the results of two valleys with $\tau_z=1$ and
$\tau_z=-1$, respectively, and in the cyan region the Hall conductance vanishes.
(e) and (f) Zoomed in view of (c) and (d), respectively.
Here, $B_0=m_z^2/(e\hbar v^2)$. }
\label{Fig2}
\end{figure}

Without $m_z$, energy reads $E_m=0,\pm E_B\sqrt{2m+1}$. Different from a single zero energy LL in a single layer graphene and two zero energy levels in bilayer graphene, there are infinite zero energy LLs in the massless spin-1 system (each case when $m=-1$ or $m>0$ will contribute a zero energy level).
However, these zero LLs do not contribute to the Hall conductance in contrast to graphene systems~\cite{Lan2011}.
In addition, the energy relation is
distinct from $\pm E_B\sqrt{m}$ in a single layer graphene and $\pm E_B\sqrt{m(m-1)}$ in bilayer graphene. With $m_z$,
the eigenenergy in a single layer graphene and bilayer graphene reads $\pm E_B\sqrt{m+m_z^2}$ and $\pm E_B\sqrt{m(m-1)+m_z^2}$,
respectively, implying that the zero energy level is gapped with energy $m_z$. In our system,
for $m=-1$, zero energy becomes $-m_z$, and for $m>0$ new energy around zero can be
approximated by $m_z/(2m+1)$, which is the first order correction when
$m_z$ is relatively small. This indicates that the zero energy LLs develop into a series of fine-structure levels in the
presence of a mass term.

Before displaying LL structures in detail, it would be helpful to discuss LLs from the semiclassical quantization rule~\cite{Niu1999,Hatsugai2009}:
\begin{equation}
A(E_m)=(m+1/2-\Gamma/2\pi)2\pi eB/\hbar,
\end{equation}
where $A(E_m)$ is the momentum space area enclosed by the closed cyclotron
orbit and $\Gamma$ is the Berry phase along the orbit. For massless spin-1 fermions,
we are able to obtain the LLs by the above formula while setting $\Gamma=0$. For massive fermions, however,
since cyclotron orbits do not exist inside the gap, fine-structure levels can not be predicted
by the semiclassical theory.

In Fig.~2(c) and (d), we plot energy of LLs of the Hamiltonian (\ref{HM}) with respect to
the magnetic field strength $B$ for fixed $m_z$ and to $m_z$ for fixed $B$, respectively.
In (c), we see that the levels can be divided into two groups: one corresponding to the states with the absolute
eigenenergy increasing with $\sqrt{B/B_0}$ and the other to those with constant energy for large
$B$. In the latter group, there are two states with exact constant energy equal to $\pm m_z$ and others split from zero energy
and reaching a constant value $m_z/(2m+1)$ for large $B$. Interestingly, the state with energy $\pm m_z$ crosses
with another state with energy $\pm(-m_z+\sqrt{m_z^2+4E_B^2})/2$ for the other valley at $B=2B_0$.
On the other hand, the band structure of LLs as a function of $m_z$ in Fig.~\ref{Fig2}(d) explicitly demonstrates
that $m_z$ will break the degeneracy of two valleys and lift the infinite degeneracy of zero energy LLs,
leading to a series of bands around zero. The crossing between states with energy $\pm m_z$ and
$\pm(-m_z+\sqrt{m_z^2+4E_B^2})/2$ for the other valley is also manifested in this figure, which
occurs at $m_z=E_B/\sqrt{2}$. In both figures, we also explicitly display the region
where the Hall conductance vanishes; it increases (decreases) by one as the Fermi surface moves up (down) across a
LL. For clarity, we plot the zoomed in figures of Fig.~\ref{Fig2}(c) and (d) in Fig.~\ref{Fig2}(e) and (f),
clearly showing the emergence of a series of LLs around zero. In Fig.~\ref{Fig2}(e), as we raise the
magnetic field following the upper (lower) arrow, we will see the change of the Hall conductance from 0 to $e^2/h$
(from 0 to $e^2/h$ and finally to $2e^2/h$), suggesting an experimental signature in terms of magnetic fields.

\begin{figure}[t]
\includegraphics[width=3.2in]{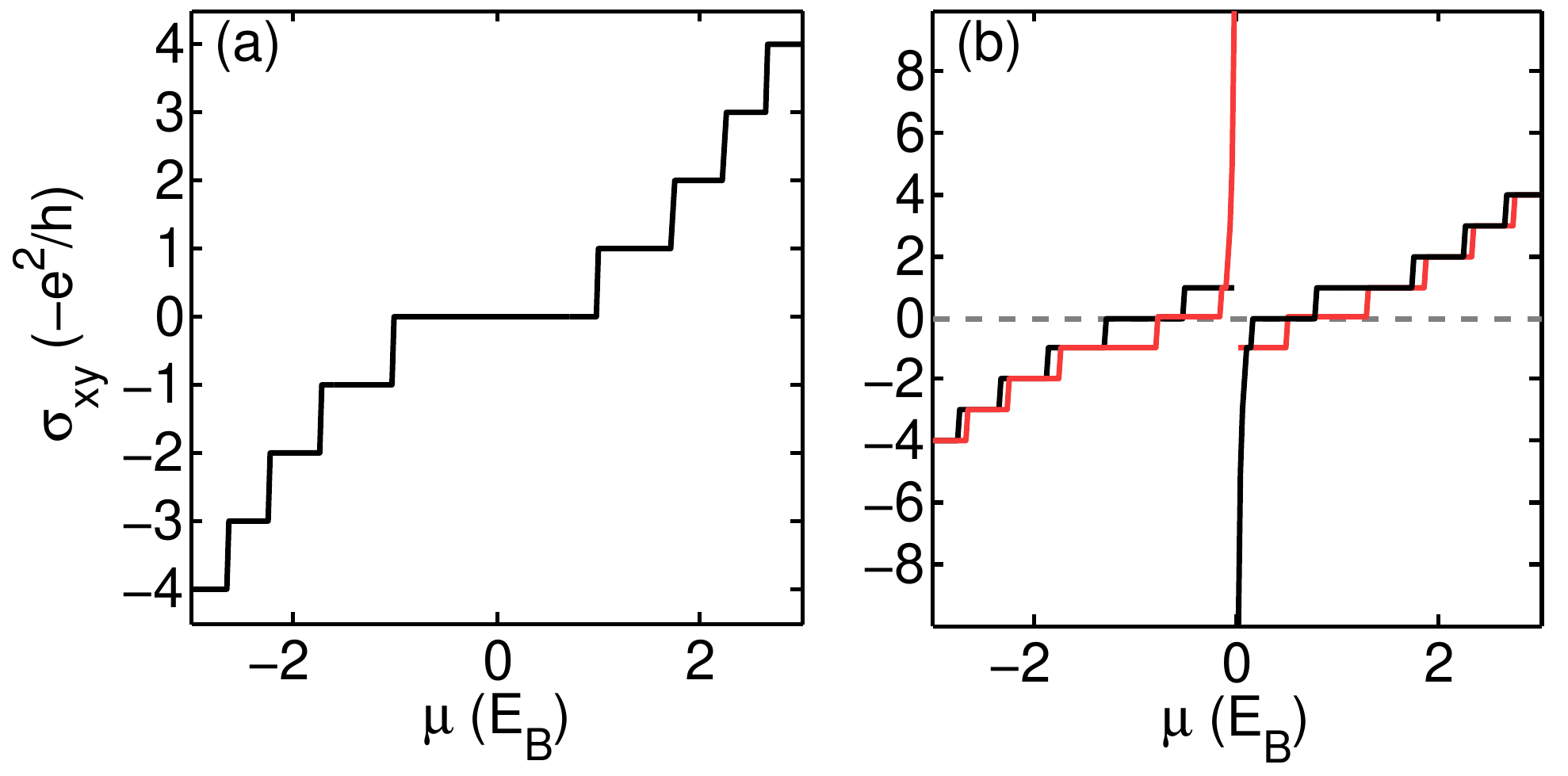}
\caption{(Color online)
The Hall conductance of a single valley for massless electrons with
$m_z=0$ in (a) and massive ones with $m_z=0.5E_B$ in (b). The black and red lines denote the
valley with $\tau_z=1$ and $\tau_z=-1$, respectively.}
\label{Fig3}
\end{figure}

To calculate the Hall conductance of the system at zero temperature, we employ the Kubo formula~\cite{Nicol2014},
\begin{equation}
\sigma_{xy}=\frac{i\hbar e^2}{\Omega}\sum_{\substack{E_a<E_F \\ E_b\geqslant E_F}}\frac{1}{(E_b-E_a)^2}(j_x^{ab}j_y^{ba}-j_y^{ab}j_x^{ba}),
\end{equation}
where $E_a$ and $E_b$ are eigenenergy of eigenstates $|a\rangle$ and $|b\rangle$, respectively,
$j_\nu^{ab}=\langle a|\partial H_M/\partial p_\nu |b\rangle$ with $\nu=x,y$ are current matrix elements
between these eigenstates, $E_F$ is the Fermi energy, and $\Omega$ is the area of the system.

In Fig.~\ref{Fig1}(d), we plot the Hall conductance of our system calculated by the Kubo formula
with respect to the chemical potential. For massless electrons without $m_z$, the conductance (denoted by the red line)
can take only even numbers due to
the degeneracy of two valleys' LLs, but drops to zero as the chemical potential is moved into the gap between zero energy and
the first nonzero energy LLs, reminiscent of massive bilayer graphene.
For massive electrons with $m_z$, since the degeneracy of two valleys is broken, the conductance can take both odd and even numbers.
Moreover, with the decline of the chemical potential from the positive energy, the conductance first
decreases to zero and then rises dramatically in the opposite direction as an infinite ladder of fine staircase; with the further decline
of the chemical potential across zero, the conductance suddenly flips its sign due to a
Van Hove singularity~\cite{Hatsugai2006} at zero energy. The same phenomenon also
occurs for a negative chemical potential owing to the antisymmetric Hall conductance with respect to
zero energy.

\begin{figure*}[t]
\includegraphics[width=\textwidth]{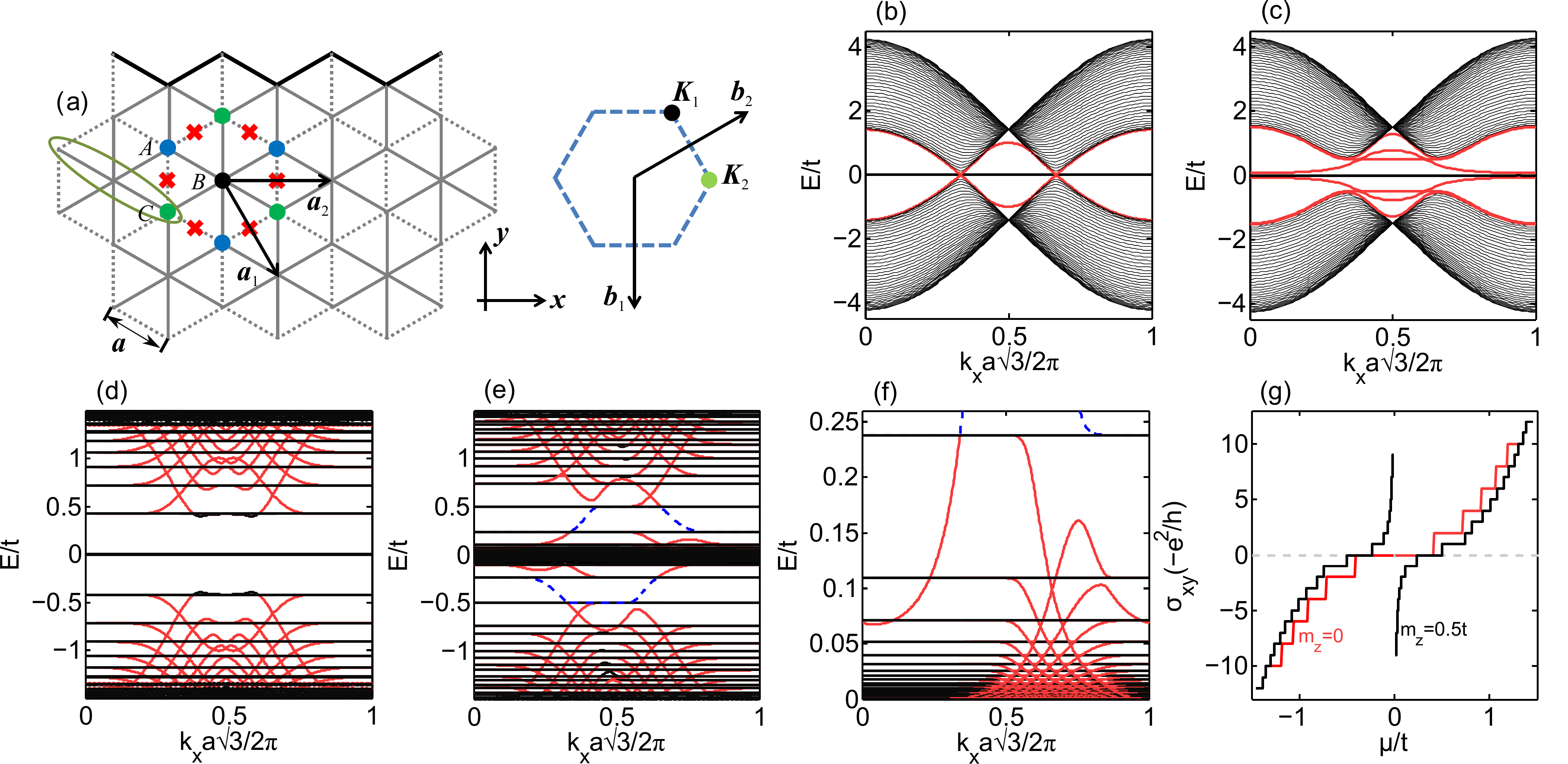}
\caption{(Color online) (a) Schematics of a lattice structure of a tight-binding model with
a pair of triply degenerate points. Each unit cell consists of three sites: A, B and C, represented
by cyan, black and green solid circles, respectively. There is no hopping between A and
C sublattices.
The black line shows the zigzag boundary condition. The right one in (a)
displays the Brillourn zone with ${\bf b}_1$ and ${\bf b}_2$ being the reciprocal vectors.
Band structure of our tight-binding model subject to zigzag boundary conditions without
magnetic fields in (b) with $m_z=0$ and in (c) with $m_z=0.5t$, and with magnetic fields
in (d) with $m_z=0$ and in (e) with $m_z=0.5t$, where the red lines
denote the edge states. (f) is the zoomed in view of (e) around zero energy and
(g) plots the corresponding Hall conductance. In (d-f), the flat black lines represent
the LLs. The dashed blue lines in (d-f) are associated with the states located
at the same edge and hence not chiral. Here, we use 240 unit cells with the magnetic flux per
unit cell being $1/60 \phi_0$.}
\label{Fig4}
\end{figure*}

To see the Hall conductance of each single valley, we plot them in Fig.~\ref{Fig3}.
For massless electrons without $m_z$, two valleys exhibit the same integer Hall conductance due to the degeneracy of
their LLs, giving rise to the total even number Hall conductance. For massive electrons with $m_z$, the degeneracy
is broken, and their Hall conductances are distinct and antisymmetric with respect to
zero energy. For the valley with $\tau_z=1$ (labeled by the black line), as we decrease the chemical potential
from positive values toward zero, the Hall conductance first drops to zero and then revives in the opposite
direction as an infinite ladder of fine staircase with the chemical potential moved into the fine-structure LLs.
As the chemical potential
is further moved across zero, the conductance (in units of $-e^2/h$) suddenly changes to 1 and then decreases. For the other valley, the
same phenomenon occurs but in an antisymmetric manner with respect to zero energy. We note that no fractional
Hall conductance appears, which is in sharp contrast to a single Dirac cone that can have a half Hall conductance.

\section{Dice model}

We consider the dice model~\cite{Sutherland1986} to realize this peculiar QHE. In the model visualized in Fig.~\ref{Fig4}(a),
a sublattice denoted by B is added in the center of a regular hexagon in a hexagonal lattice with the assumption that particles
on the neighbor sites A and C can hop only to site B, but not to each other, which
may occur if there exist high energy barriers between them~\cite{Rizzi2006}. Let us focus on the spinless case
and write down the tight-binding Hamiltonian in the real space,
\begin{equation}
H_{dice}=(-t\sum_{\langle i,j\rangle}{\hat{B}_i^\dagger \hat{A}_j}-
t\sum_{\langle i,j\rangle}{\hat{B}_i^\dagger \hat{C}_{j}}+H.c.)+H_z,
\end{equation}
where $\hat{D}_i^\dagger$ ($\hat{D}_j$) with $D=A,B,C$ creates (annihilates) a particle on sublattice D of
site ${\bf R}_i$, $\langle\rangle$ represents the nearest-neighbor sites and $t$ the tunneling
strength, and $H_z=m_z\sum_i(\hat{A}^\dagger_i\hat{A}_i-\hat{C}^\dagger_i\hat{C}_i)$ is the mass term.
Using periodic boundary conditions, we can write down the Hamiltonian in the momentum space,
\begin{equation}
H_{dice}({\bf k})=\left(
             \begin{array}{ccc}
               m_z & d({\bf k}) & 0 \\
               d({\bf k})^* & 0 & d({\bf k}) \\
               0 & d({\bf k})^* & -m_z \\
             \end{array}
           \right),
\end{equation}
where $d({\bf k})=1+e^{-i{\bf k}\cdot {\bm a}_1}+e^{-i{\bf k}\cdot {\bm a}_2}$ with ${\bf k}$ being
the momentum and ${\bm a}_1$ and ${\bm a}_2$ being primitive vectors. Near the ${\bf K}$
(${\bf K}^\prime$) point in the Brillouin zone, the Hamiltonian is approximated by
$H_{TB}({\bf q})=v_0[\mp (q_x+\sqrt{3}q_y) \hat{S}_x+(-\sqrt{3}q_x+q_y)\hat{S}_y]+m_z\hat{S}_z$ with
$v_0=3a/4$ and $a$ being the lattice constant, where ${\bf q}$ is measured with respect to ${\bf K}_1$ (${\bf K}_2$),
thereby demonstrating the emergence of two triply degenerate points located at ${\bf K}_1$ and ${\bf K}_2$
when $m_z=0$.

\begin{figure*}[t]
\includegraphics[width=5.5in]{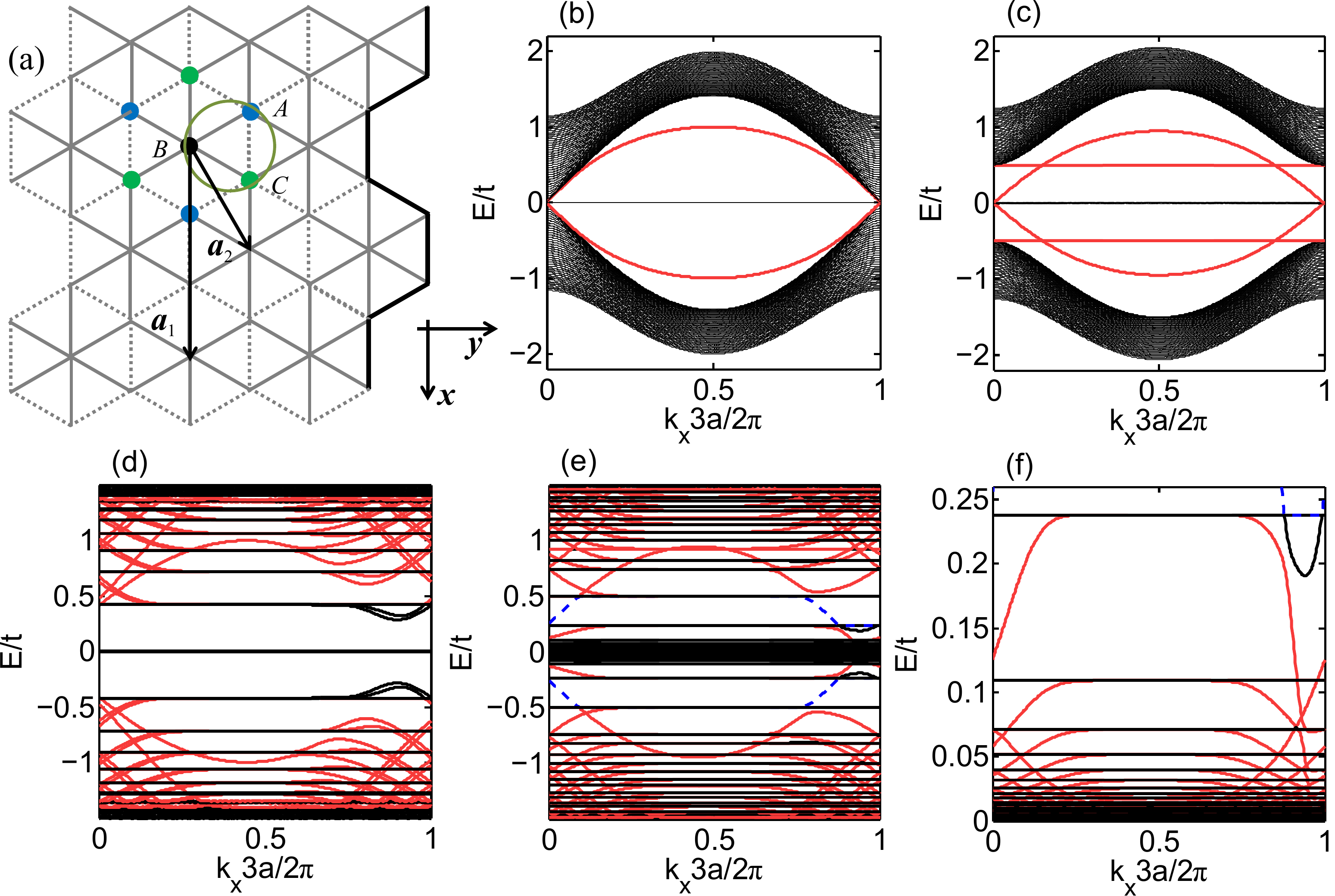}
\caption{(Color online) (a) Schematics of a lattice structure with armchair boundaries denoted by
the black line. (b-f) The band structure obtained by using the same parameters as those in Fig.~\ref{Fig4}(b-f),
but with armchair boundary conditions.}
\label{Fig5}
\end{figure*}

In a geometry with zigzag edges
in the $y$ direction and periodic boundary conditions in the $x$ direction, we calculate the band
structure of the tight-binding model and plot them in Fig.~\ref{Fig4}(b) and (c), corresponding to
the case without and with $m_z$, respectively. In (c) with $m_z$, we see that there emerge
three particle and three valence edge state bands denoted by the red lines: Two of particle (hole)
bands have dispersion and the other one connecting two energy minima (maxima) is
flat and dispersionless. When the band gap decreases, one particle (hole) edge bands shrink into zero
energy and completely mix with the zero energy bulk state when the band gap vanishes, as shown in
Fig.~\ref{Fig4}(b). Additionally, the other two particle (hole) edge bands become degenerate;
this degeneracy only exists for zigzag boundaries and does not hold under armchair boundary
conditions. For the latter, we present the band structure in Fig.~\ref{Fig5}(b) and (c), showing that
there are two particle (hole) edge states with one being dispersionless for massive electrons, which will mix
into the zero energy bulk states for massless electrons. In the massless case, the particle (hole) edge
state with dispersion is not degenerate, different from the zigzag scenario.

In the presence of magnetic fields along the $z$ direction, if choosing a unit cell as shown in Fig.~\ref{Fig4}(a),
we write down the Hamiltonian using the Peierls substitution
\begin{widetext}
\begin{eqnarray}
H_M=&&-t\sum_{m,n}\left[e^{i\alpha(-m+1/6)\pi}\hat{A}^\dagger_{m,n}\hat{B}_{m,n}+e^{i\alpha(m+1/6)\pi}\hat{C}^\dagger_{m,n}\hat{B}_{m,n}
+e^{-i\alpha(m+1/6)\pi}\hat{C}^\dagger_{m,n-1}\hat{B}_{m,n} \right. \nonumber \\
&&\left.+e^{i\alpha(m-1/6)\pi}\hat{A}^\dagger_{m,n+1}\hat{B}_{m,n}
+\hat{C}^\dagger_{m-1,n}\hat{B}_{m,n}+\hat{A}^\dagger_{m+1,n}\hat{B}_{m,n}+H.c.\right]+H_z
\end{eqnarray}
\end{widetext}
in the Landau gauge: $A_x=0$ and $A_y=B x$, where $\alpha=\phi/\phi_0$ with $\phi_0=h/e$ and $\phi$ being the magnetic flux per unit cell.

Because the quantized Hall conductance is equal to the number of chiral edge states
originated from the topological property of LLs, we can visualize the Hall conductance
by displaying the energy spectra under open boundary conditions. For zigzag boundaries,
the spectra is shown in Fig.~\ref{Fig4}(d-f),
where the black and red lines represent the LLs and edge states, respectively.
We see from (d) that there emerge even
numbers of chiral edge states for massless electrons with $m_z=0$ in the gaps except the first gap,
where no chiral edge states exist. This implies that the Hall conductance takes even numbers
and becomes zero when the Fermi surface lies in the first gap as shown in Fig.~\ref{Fig4}(g),
which is in agreement with our continuous model's results. It is different from a spinless single
layer graphene and bilayer graphene, where there exist one and two chiral edge states in the first
gap~\cite{Neto2009,Abergel2010}, respectively. We see the same chiral edge state behavior under armchair
boundary conditions in Fig.~\ref{Fig5}(d).

For massive electrons with $m_z$, the double degeneracy of each Landau level with nonzero energy due
to the existence of two valleys is lifted, so that the number of chiral edge states can increase (decrease) by one
with the change of the chemical potential as shown in Fig.~\ref{Fig4}(e). This suggests that the Hall
conductance can take both even and odd integer values as shown in Fig.~\ref{Fig4}(g). Additionally, the LLs with zero energy
develop into a series of bands around zero. In the gaps between these LLs, remarkably,
there appear chiral edge states and their number increases as energy moves toward zero, giving rise to the
revival of QHE as illustrated in Fig.~\ref{Fig4}(g). Since the chirality of the edge states with positive (negative)
energy around zero is opposite to that of the other states with positive (negative) energy,
the Hall conductance changes its sign during the revival. Furthermore, the chirality of
the edge states flips its sign as energy moves across zero, the Hall conductance will accordingly experience
a sudden sign reversal during the process because of a Van Hove singularity at zero energy.
All these results agree well with our continuous model's results. We see the same chiral edge state behavior under armchair
boundary conditions in Fig.~\ref{Fig5}(e) and (f).

Furthermore, the Hall conductance in the lattice model can be calculated by the formula
\begin{equation}
\sigma_{xy}=-\frac{e^2}{h}C_F(E_F),
\end{equation}
where $C_F(E_F)=\sum_{E_n<E_F}C_n$ with $C_n$ denoting the Chern number of the LL with energy $E_n$. We employ the method
proposed in Ref.\cite{Hatsugai2005} to compute $C_F(E_F)$ in the discretized Brillouin zone
meshed as $\{ {\bf k}_l\}$~\cite{Hatsugai2005,Hatsugai2009},
\begin{equation}
C_F=\frac{1}{2\pi i}\sum_{l}F_{12}({\bf{k}}_l),
\end{equation}
where
\begin{gather}
F_{12}({\bf k})=\text{ln}\left[D_1({\bf k})D_2({\bf k}+\delta {\bf k}_1)/D_1({\bf k}+\delta {\bf k}_2)D_2({\bf k})\right], \\
D_\mu({\bf k})=\text{det}U_\mu({\bf k})/|\text{det}U_\mu({\bf k})|,\\
U_\mu({\bf k})=\psi^\dagger({\bf k})\psi({\bf k}+\delta{\bf k}_\mu),\\
\psi({\bf k})=[\begin{array}{cccc}
                     u_1({\bf k}) & u_2({\bf k}) & \ldots & u_N({\bf k})
                   \end{array}
],
\end{gather}
with $\delta {\bf k}_\nu=|\delta {\bf k}_\nu|{\bf b}_\nu/|{\bf b}_\nu|$ ($\nu=1,2$), $u_n({\bf k})$ being the \emph{n}th eigenstate of $H_M$ and $u_N$ being the highest eigenstate below the Fermi surface.

We perform the calculation of the Hall conductance in the dice model using the above approach and plot it in Fig.~\ref{Fig4}(g).
It is evident that the Hall conductance takes the same value as the number of chiral edge states as shown in Fig.~\ref{Fig4}(d-f)
and Fig.~\ref{Fig5}(d-f).

\section{conclusion}

We have studied QHEs in a 2D massive spin-1 fermion system and found that the Hall conductance first revives as an infinite ladder of fine staircase after it crosses the zero plateau when the chemical potential is moved toward zero energy and then suddenly reverses
with its sign being flipped when the chemical potential is moved across zero, in sharp contrast to
the conventional QHE in normal materials or the unconventional QHE for Dirac fermions. Although the sudden jump of the Hall conductance also happens in lattice models, e.g. graphene, at high energy, our results show that this phenomenon occurs in a continuous massive spin-1 fermion model around zero energy.
Moreover, we have explored the peculiar QHEs in a gapped dice model and found the same phenomena as the continuous spin-1 fermion model.
The dice model may be realized in cold atom systems with an appropriate potential being generated by lasers~\cite{Rizzi2006} and an artificial magnetic fields being engineered by the laser-assisted tunneling technology that has implemented strong magnetic fields in square optical lattices~\cite{Bloch2013,Kettele2013}.

\begin{acknowledgments}
We thank S.-T Wang, Y.-H. Zhang, F. Zhang, D.-W. Zhang, S. A. Yang and T. Biswas for helpful discussions. This work was supported by the ARL, the IARPA LogiQ program, and the AFOSR MURI program.
\end{acknowledgments}

\end{document}